\documentclass[amsmath,
twocolumn,
aps,prl]{revtex4}
\usepackage{graphicx}% Include figure files
\usepackage{dcolumn}% Align table columns on decimal point
\usepackage[colorlinks=true, pdfstartview=FitV, linkcolor=blue, 
            citecolor=blue, urlcolor=blue]{hyperref}

\begin{document}

\title{An off-board quantum point contact as a sensitive detector of
  cantilever motion} \author{M. Poggio$^{1,2}$, M. P. Jura$^3$, C. L.
  Degen$^1$, M. A. Topinka$^{4,5}$, H. J.  Mamin$^1$, D.
  Goldhaber-Gordon$^4$ and D.  Rugar$^1$} \affiliation{$^1$IBM
  Research Division, Almaden Research Center, 650 Harry Rd., San Jose
  CA, 95120 \\ $^2$Center for Probing the Nanoscale, Stanford
  University, Stanford, CA 94305 \\ $^3$Department of
  Applied Physics, Stanford University, Stanford, CA 94305 \\
  $^4$Department of Physics, Stanford University, Stanford, CA 94305
  \\ $^5$Department of Material Science and Engineering, Stanford
  University, Stanford, CA 94305} \date{\today}

\maketitle

\textbf{ Recent advances in the fabrication of microelectromechanical
  systems (MEMS) and their evolution into nanoelectromechanical
  systems (NEMS) have allowed researchers to measure extremely small
  forces, masses, and displacements \cite{Schwab:2005}.  In
  particular, researchers have developed position transducers with
  resolution approaching the uncertainty limit set by quantum
  mechanics \cite{Knobel:2003,LaHaye:2004,Naik:2006,Flowers:2007}.
  The achievement of such resolution has implications not only for the
  detection of quantum behavior in mechanical systems, but also for a
  variety of other precision experiments including the bounding of
  deviations from Newtonian gravity at short distances
  \cite{Smullin:2005} and the measurement of single spins
  \cite{Rugar:2004}.  Here we demonstrate the use of a quantum point
  contact (QPC) as a sensitive displacement detector capable of
  sensing the low-temperature thermal motion of a nearby
  micromechanical cantilever.  Advantages of this approach include
  versatility due to its off-board design, compatibility with
  nanoscale oscillators, and, with further development, the potential
  to achieve quantum limited displacement detection
  \cite{Clerk:2003}.}

At present, the most sensitive displacement detectors for nanoscale
mechanical resonators rely on the mechanical modulation of current
flow through a single electron transistor or atomic point contact,
achieving resolutions around $10^{-15}$ m/$\sqrt{\text{Hz}}$, which in
one case is only several times the quantum limit
\cite{Knobel:2003,Naik:2006,LaHaye:2004,Flowers:2007}.  These devices,
however, feature a resonator and sensor integrated into a single unit,
limiting their versatility for some force sensing applications.  While
high finesse interferometers also achieve nearly quantum limited
displacement resolution --- down to an astounding $10^{-20}$
m/$\sqrt{\text{Hz}}$ \cite{Caniard:2007,Arcizet:2006} --- their
application to micro- and nanomechanical oscillators is challenging,
especially as oscillator size is reduced.  A fundamental obstacle is
the optical diffraction limit, which sets a rough lower bound on the
size of the measured oscillator.  In addition, many requirements of a
high finesse cavity (e.g.\ thick substrates and stiff multilayer
mirror stacks for maximum reflection) run counter to the requirements
of the most sensitive MEMS and NEMS (e.g.\ low spring constants and
thin membranes for sensitive force detection).  Optical
interferometers encounter some limitations at the low temperatures
often required in ultra-sensitive force microscopy.  For temperatures
below 1 K, the absorption of light from a typical interferometer laser
--- even for incident powers less than 100 nW --- has been observed to
heat Si cantilevers through absorption
\cite{Bleszynski:2007,Mamin:2001}.  As a result, the resolution of
typical optical interferometry of micromechanical force sensors hovers
above $10^{-13}$ m/$\sqrt{\text{Hz}}$
\cite{Nonnenmacher:1992,Mamin:2001}.

In this paper we show how simply bringing a micromechanical oscillator
in close proximity to an off-board QPC allows for sensitive
displacement measurements.  In demonstrating this principle, our QPC
detector achieves a resolution of $10^{-12}$ m/$\sqrt{\text{Hz}}$,
which is comparable to that achieved by optical interferometry on
resonators of similar size.  Our QPC transducer, however, has the
fundamental advantage that it can be applied to measurements of
oscillators with dimensions smaller than the optical diffraction
limit.  In addition, by virtue of its off-board design, the QPC can be
used in conjunction with sensitive cantilevers in a variety of force
sensing applications including magnetic resonance force microscopy
\cite{Sidles:1995}.  While the resolution of our QPC is limited by
device imperfections, QPC transducers of this type have the properties
required to reach the quantum limit on continuous position detection
\cite{Clerk:2003}.

\begin{figure*}[t]\includegraphics{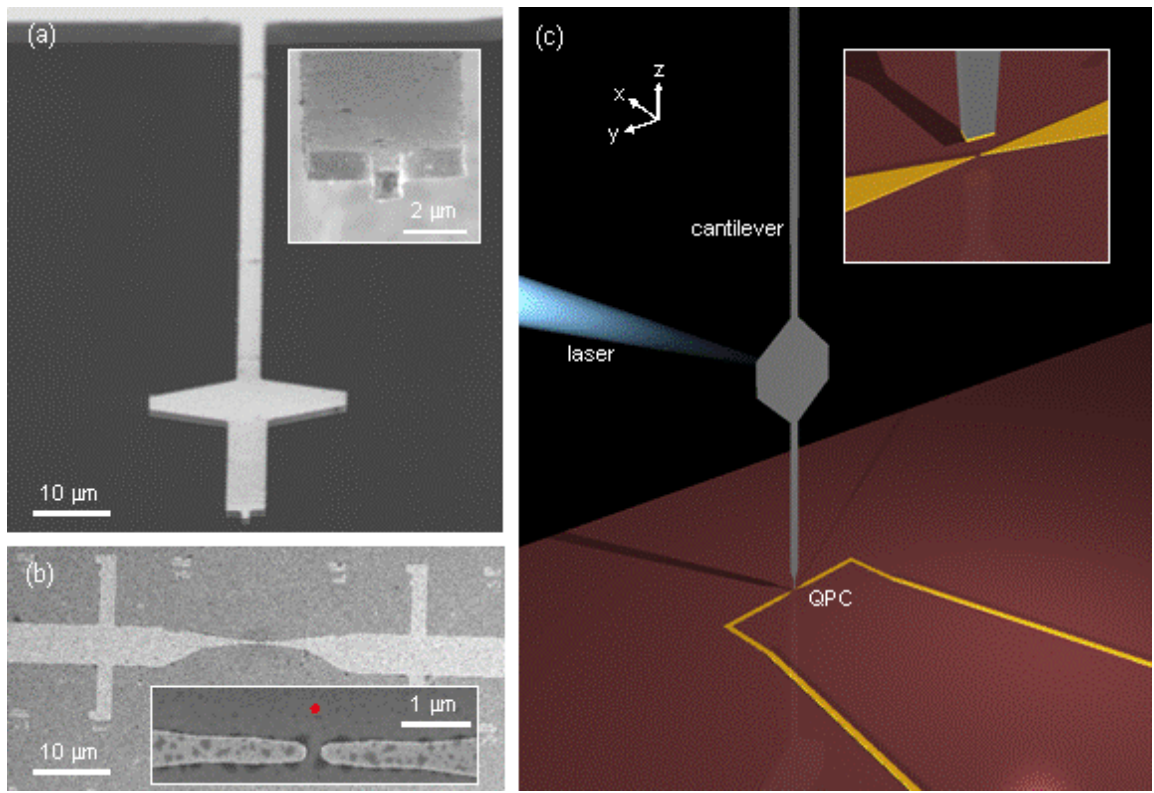}
\caption{\label{fig1}
  (a) Scanning electron micrograph (SEM) of the cantilever and its
  Au-coated tip (inset).  (b) SEM of the QPC with high-resolution view
  of the active region (inset).  A red dot indicates the position of
  the cantilever tip during the experiment.  (c) Scaled schematic of
  the experimental set-up.  A close-up view of the QPC -- with the
  cantilever in close proximity -- is shown in the inset.  The laser
  beam is part of the low-power interferometer used to calibrate
  displacement measurements made by the QPC.}\end{figure*}

In the years since the discovery of quantized conductance through
semiconductor QPCs \cite{vanWees:1988,Wharam:1988}, these devices have
been used as sensitive charge detectors in a variety of applications.
The dependence of a QPC's source-drain conductance on small changes in
electrostatic fields make it useful as a detector of single electrons
in gate-defined quantum dots (QDs) \cite{Field:1993} or of charge
motion through electron interferometers
\cite{Buks:1998,Sprinzak:2000}.  This extreme sensitivity to charge
has also been applied in the detection of mechanical motion on minute
scales.  In 2002, Cleland \textit{et al}.\ demonstrated a displacement
detector with a resolution of $3 \times 10^{-12}$ m/$\sqrt{\text{Hz}}$
at 1.5 MHz using the piezoelectric effect in a GaAs micromechanical
resonator to modulate current through an integrated QPC
\cite{Cleland:2002}.  Since this method requires the QPC to be built
into a piezoelectric resonator, device processing can degrade both the
mobility of the two-dimensional electron gas (2DEG) forming the QPC
and the quality factor of the resonator.  Furthermore, the stiff
doubly-clamped geometry of the resonator and the requirement that it
be made from a single crystal GaAs heterostructure limit its
application as sensitive force detector.  Here we use a different
scheme for displacement detection wherein a cantilever is brought
close to an off-board QPC causing the lever's motion to modulate the
QPC conductance.  In principle, the motion of an arbitrary resonator,
without any integrated devices, can be detected in this way.

The displacement measurement, carried out in vacuum (pressure $< 1
\times 10^{-6}$ torr) at $T = 4.2$ K, is made by positioning the tip
of a metal-coated Si cantilever about 100 nm above a QPC, as shown
schematically in Fig.~\ref{fig1}(c).  Due to the tip's proximity to
the QPC itself --- i.e.\ the narrow channel of electron conduction
directly between the gates --- the lever tip and the QPC are
capacitively coupled.  The tip acts as a movable third gate above the
device surface.  Changes in the cantilever potential $V_l$ affect the
potential landscape of the QPC channel and thereby alter its
conductance $G$.  A voltage $V_g$ applied to the two gates patterned
on the surface modifies $G$ in the same manner.

\begin{figure}[t]\includegraphics{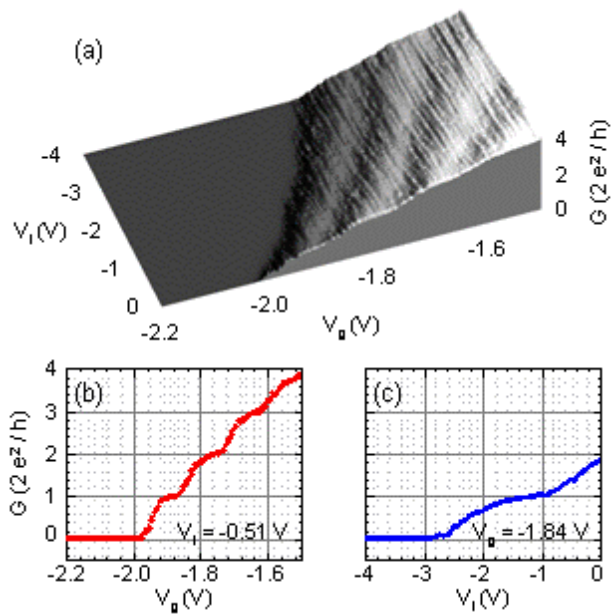}
\caption{\label{fig2}
  (a) Conductance plotted as a function of gate voltage and lever
  voltage with the cantilever tip positioned as shown in the inset to
  Fig.~\ref{fig1}(b) and at $z = 70$ nm above the QPC.  (b) and (c)
  show linecuts of (a) for constant $V_l$ and $V_g$ respectively.
}\end{figure}

Fig. \ref{fig2} shows the dependence of $G$ on both $V_g$ and $V_l$
with the cantilever positioned near the QPC.  The tip is located $z =
70$ nm above the surface of the QPC device and $x = 660$ nm directly
in front of the point contact, as indicated in the inset to
Fig.~\ref{fig1}(b).  As $V_g$ and $V_l$ are made more negative, both
act to decrease $G$ in steps of the conductance quantum $2 e^2 / h$
until the conductance through the point contact pinches off.  From
Figs.~\ref{fig2}(a) and (b) we determine that $G$ is about 14 times
more sensitive to changes in $V_g$ than to changes in $V_l$.  This
factor corresponds to the ratio between the gate-QPC capacitance and
the tip-QPC capacitance, $C_g/C_l$.

\begin{figure}[t]\includegraphics{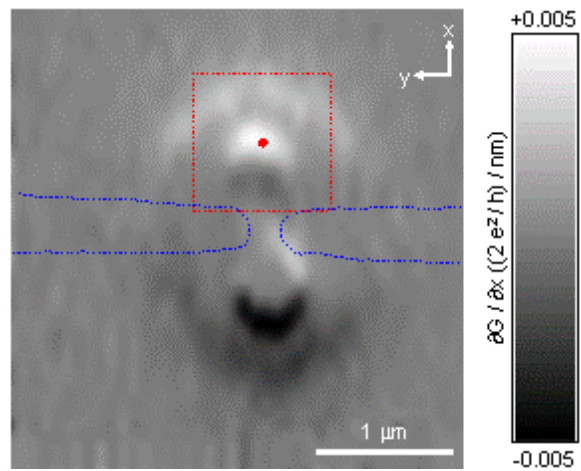}
\caption{\label{fig3}
  $\partial G / \partial x$ plotted as a function of cantilever $x$
  and $y$ over the QPC device.  Blue dotted lines show the position of
  the QPC gates, while a red dot and red dotted lines indicate the
  extent of the cantilever tip and its position during the other
  measurements.  $z = 70$ nm, $V_l = -3.0$ V, and $V_g = -1.75$ V.
}\end{figure}

The tip-QPC capacitive coupling depends strongly on their relative
separation; only when the tip is positioned near the QPC, does $V_l$
affect $G$ strongly.  By moving the tip over the device surface at
fixed distance $z$ and with a voltage $V_l$ applied, we can make an
image of its effect on $G$ and map this capacitive coupling.  In
regions near the point contact, where changes in the lever position
most strongly affect $G$, we find a conductance response of up to
0.005 $(2 e^2 / h)$/nm of cantilever motion along $x$ as shown in
Fig.~\ref{fig3}.  The position of the cantilever tip during the
following experiments, indicated again in Fig.~\ref{fig3}, corresponds
to a region where $G$ responds most sensitively to changes in lever
position.

\begin{figure}[t]\includegraphics{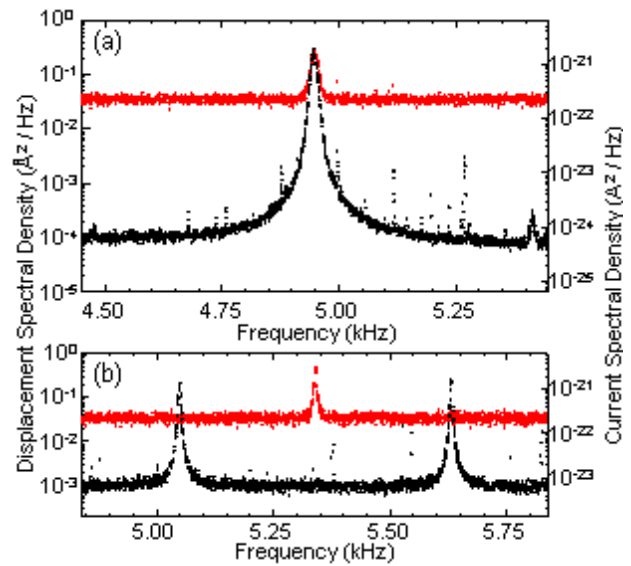}
\caption{\label{fig4}
  Cantilever thermal noise spectrum observed using a QPC transducer.
  (a) QPC with DC source-drain bias.  The thermal motion of the
  cantilever tip, measured by an optical interferometer, is plotted in
  red in \AA$^2$/Hz.  The tip is $z = 70$ nm above the QPC.  The
  response to this motion by the QPC is plotted in black for $V_{sd} =
  2.0$ mV DC, $V_g = -1.74$ V, and $V_l = -3.0$ V in both $A^2$/Hz and
  \AA$^2$/Hz.  (b) QPC with AC source-drain bias.  The same
  description applies as in (a) except that we apply a 2.0 mV$_{rms}$
  sinusoid at 291 Hz to $V_{sd}$ with $V_g = -1.45$ V and $V_l = -3.0$
  V \cite{footnote:2}.  }\end{figure}

With the cantilever so positioned, we study the QPC's effectiveness as
a transducer of tip motion.  Fig.~\ref{fig4} shows the displacement
resolution of the QPC as compared with a low-power laser
interferometer simultaneously detecting the cantilever's average
thermal motion.  At $T = 4.2$ K, the cantilever tip has a thermal
motion amplitude of $x_{th} = 1.6$ \AA$_{rms}$.  In Fig.~\ref{fig4}(a)
we plot in red the spectral density of the cantilever displacement
$S_x$ driven by thermal force noise as measured by the optical
interferometer.  In black we plot the spectral density $S_i$ of the
current driven through the point contact with a DC source-drain
voltage $V_{sd} = 2.0$ mV, $V_g = -1.74$ V, and $V_l = -3.0$ V.  The
current response of the QPC matches the cantilever thermal noise in
both frequency and quality factor.  Once we normalize the peak QPC
response to the peak amplitude of cantilever motion as measured by the
interferometer, we calculate a conductance sensitivity of 0.004 $(2
e^2 / h)$/nm of cantilever motion.  Furthermore, this normalization
allows us to plot the spectral density of the QPC response in
Fig.~\ref{fig4} both in terms of current on the right axis and in
terms of cantilever motion on the left axis.  The level of the noise
floor sets the resolution of the QPC displacement transducer at
$10^{-12}$ m/$\sqrt{\text{Hz}}$, which is over an order of magnitude
better than the low-power optical interferometer used here.  In our
experiment, the noise on the QPC current and thus the displacement
resolution of the transducer is limited by charging noise (charges
fluctuating near the QPC) with a $1/f$ dependence \cite{Li:1990}.

In order to verify that the QPC response at the cantilever resonant
frequency $\nu_c$ is not produced by electrical feed-through, we drive
the QPC with an AC source-drain voltage $V_{sd} = 2.0$ mV$_{rms}$ at
291 Hz.  In Fig.~\ref{fig4}(b) the spectrum of the source-drain
current through the QPC reveals a response centered on $\nu_c$ and
split between two peaks spaced by twice the source-drain drive
frequency.  These sidebands are the signature of a mixer and they
confirm that the response of the QPC results from the thermal motion
of the nearby cantilever.  The cantilever tip indeed acts as an
oscillating gate which modulates the QPC conductance.  Further
evidence comes from exciting cantilever oscillations using a
mechanically coupled piezoelectric element: the sideband amplitude
increases as a function of increasing excitation amplitude.

In order to extract parameters such as the cantilever quality factor
$Q$ and average thermal motion $x_{th}$, we fit the spectral density
of the cantilever displacement as measured by the optical
interferometer to the expected (single-sided) spectrum from a simple
harmonic oscillator, $S_x(\omega) = \frac{4 \omega_c^3 x_{th}^2}{Q}
\frac{1}{(\omega_c^2 - \omega^2)^2 + (\omega_c \omega / Q)^2} +
S_{x_n}$, where $\omega$ is an angular frequency, $\omega_c = 2 \pi
\nu_c$, and $S_{x_n}$ is the white spectral density of the
interferometer measurement noise.  Similarly we can quantify the
current response $i_{th}$ of the QPC displacement transducer by
fitting the spectral density of the current through the point contact
to $S_{i}(\omega) = \frac{4 \omega_c^3 i_{th}^2}{Q}
\frac{1}{(\omega_c^2 - \omega^2)^2 + (\omega_c \omega / Q)^2} +
S_{i_n}$, where $S_{i_n}$ is the white spectral density of the current
measurement noise.  We therefore define the QPC transduction factor as
$\eta = i_{th}/x_{th}$.

\begin{figure}[t]\includegraphics{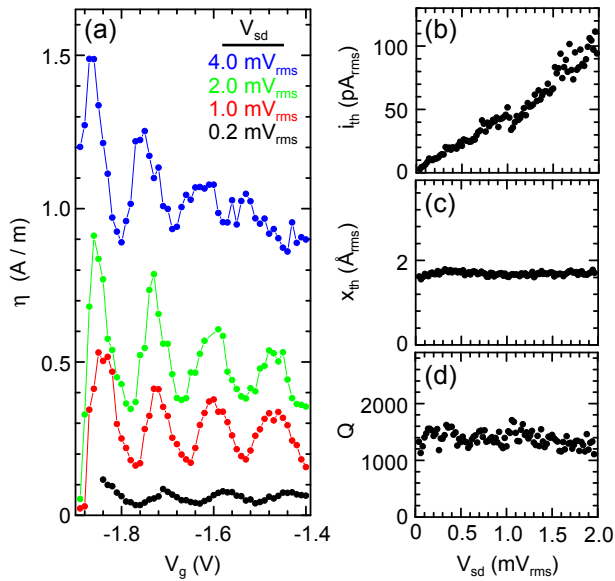}
\caption{\label{fig5}
  The response of the QPC to the thermal motion of the cantilever tip.
  The cantilever tip is $z = 70$ nm above the QPC with $V_l = -3.0$ V.
  (a) The transduction efficiency $\eta$ is plotted at different
  $V_{sd}$ as a function of $V_g$. (b) The current response $i_{th}$,
  (c) the cantilever thermal motion $x_{th}$, and (d) the cantilever
  $Q$ \cite{footnote:3} are plotted as functions of $V_{sd}$ for $V_l
  = -3.0$ V and $V_g = -1.8$ V.  }\end{figure}

In order to study the QPC's performance as a displacement transducer
and to determine optimal operating conditions, we vary several
parameters.  In Fig.~\ref{fig5}(a) we plot $\eta$ as a function of
$V_g$ for several values of $V_{sd}$ \cite{footnote:1}.  $\eta$ is
proportional to the derivative of the conductance $G$ with respect to
$V_g$ and hence shows an oscillatory behavior as a function of $V_g$.
As function of $V_g$, the maxima of $\eta$ are aligned with the steps
in $G$.  The steps --- and indeed these oscillations in $\eta$ --- are
manifestations of the quantization of conductance in QPCs.  $\eta$ has
a similar oscillatory dependence as a function of $V_l$.  In addition,
$i_{th}$ increases as a function of increasing $V_{sd}$ amplitude, as
is shown for a fixed $V_g$ in Fig.~\ref{fig5}(b).  Parameters
extracted from the interferometer spectra show that the QPC transducer
does not affect the cantilever motion.  Even for source-drain
amplitudes up to 2.0 mV$_{rms}$, both $x_{th}$ and $Q$ remain
unaffected by the QPC current within the measurement noise.

While the cantilever $Q$ is not affected by the QPC source-drain
current, it is affected by the voltages applied both to the lever and
to the QPC gates.  As has been observed before
\cite{Stipe:2001,Stowe:1997}, non-contact friction between a
cantilever and a surface degrades the cantilever's $Q$ with decreasing
tip-sample spacing.  This degradation is exacerbated by the
application of voltages to either the cantilever or the surface gates.
The observed dissipation is a result of tip-sample electric fields,
which can be present even when the lever and surface gates are
grounded \cite{Stipe:2001}.  Under typical operating conditions ($V_g
= -1.75$ V, $V_l = -3.0$ V, and $z = 70$ nm) the cantilever $Q$ is
around 2500, much lower than the intrinsic $Q_0 = 22500$ measured far
from the QPC surface.  In order to minimize this external cantilever
dissipation, future QPC transducers could be designed without surface
gates \cite{Wesstrom:1997} and without the need to apply a voltage to
the cantilever.  Our experiments show that the QPC has a substantial
response even to a grounded cantilever, possibly due to trapped
charges.  In this case ($V_l = 0$ V) with $V_{sd} = 1$ mV$_{rms}$,
$V_g = -1.98$ V, and $z = 70$ nm, we find a reasonably large $\eta$
(0.16 A/m) and much larger $Q$ (7500) than with a voltage applied to
the lever.

The prospects for improving the resolution of future QPC transducers
based on this proof of principle are bright.  By reducing the
temperature below 4.2 K, the conductance steps will become sharper,
thus increasing the response of the QPC transducer to a given
displacement.  We have also observed that lower operating temperatures
decrease the $1/f$ charging noise currently limiting our displacement
resolution.  Furthermore, the impact of this $1/f$ noise will diminish
for smaller cantilevers with higher resonant frequencies, potentially
allowing us to reach the QPC's shot noise limit.  The bandwidth of the
QPC is limited to around $10^4$ or $10^5$ Hz due to its resistance
coupled with unavoidable stray cabling capacitance.  Careful design of
both device and material parameters in order to maximize the QPC
transduction factor and minimize noise may also improve future
results.

\section{Methods}

In this experiment the cantilever and QPC are mounted in a vacuum
chamber at the bottom of a $^4$He cryostat, which is isolated from
environmental vibrations.  A three-dimensional positioning stage with
nanometer precision and stability moves the QPC relative to the
cantilever.  The QPC device is made from a heterostructure grown by
molecular beam epitaxy on a GaAs substrate.  The structure consists of
an 800 nm GaAs layer grown on top of the substrate, followed by a 15
nm Al$_{0.265}$Ga$_{0.735}$As layer, a 40 nm Si-doped
Al$_{0.265}$Ga$_{0.735}$As layer, and finally a 5 nm GaAs cap.  The
2DEG lies 60 nm below the surface with mobility $\mu = 1.0 \times
10^6$ cm$^2$ V$^{-1}$ s$^{-1}$ and a carrier density $n = 4.5 \times
10^{11}$ cm$^{-2}$ at $T = 4.2$ K, corresponding to a mean free path $
\lambda = 11$ $\mu$m.  Ti/Au (10/20 nm) gates patterned by
electron-beam lithography and shown in Fig.~\ref{fig1}(b) are used to
define the QPC within the 2DEG.  The application of a negative
potential $V_g$ between the gates and the 2DEG forms a variable width
channel through which electrons flow.  The conductance $G$ of the QPC
is measured between two ohmic contacts to the 2DEG on either side of
the channel.  A source-drain voltage $V_{sd}$ is applied across these
ohmic contacts in order to drive the conductance.

We use a metalized cantilever made from single-crystal Si that is 350
$\mu$m long, 3 $\mu$m wide, and 1 $\mu$m thick \cite{Stowe:1997}.  At
$T = 4.2$ K the cantilever has a resonant frequency $\nu_c = 5.2$ kHz
and an intrinsic quality factor $Q_0 = 22500$.  The oscillator's
spring constant is determined to be $k = 2.1$ mN/m through
measurements of its thermal noise spectrum at several different base
temperatures.  The motion of the cantilever can be detected using
laser light focused onto a 25 $\mu$m wide paddle 100 $\mu$m from its
tip and reflected back into an optical fiber interferometer
\cite{Mamin:2001}.  When the interferometer is in use, 20 nW of light
are incident on the lever from a temperature-tuned 1550-nm distributed
feedback laser diode \cite{Bruland:1999}.  As shown in
Fig.~\ref{fig1}(a), the cantilever includes a 5 $\mu$m wide section
beyond the paddle which ends with a fine 5 $\mu$m long, 1 $\mu$m wide
tip.  A thin metallic film of Cr/Au (10/30 nm), with Cr as an adhesion
layer, is evaporated onto the end of the cantilever.  22 nm of Pt
sputtered on the lever's entire surface provide a conductive path from
the cantilever base to its tip.  A voltage $V_l$ is applied to the
cantilever through a pressed-indium contact at the base of the
cantilever chip.

\begin{acknowledgments}
  We thank A. A. Clerk for useful discussions and L. S. Moore for
  measuring the properties of the 2DEG used here.  This work was
  supported by the Stanford-IBM Center for Probing the Nanoscale, an
  NSF NSEC, grant PHY-0425897.

\end{acknowledgments}


\begin{thebibliography}{28}
\bibitem{footnote:1} $\eta$ is extracted from the QPC's response to an
  AC source-drain drive voltage at 291 Hz, as shown in
  Fig.~\ref{fig4}(b).  We fit to this doubly peaked response rather
  than to the singly peaked response to a DC source-drain voltage in
  order to avoid spurious signals at $\nu_c$.  Such signals can appear
  because of small fluctuations ($< 100$ $\mu$V) in the virtual ground
  of our current amplifier.
\bibitem{footnote:2} The baseline noise of the displacement
  measurement in Fig.~\ref{fig4}(b) is somewhat worse than the noise
  in Fig.~\ref{fig4}(a) because a smaller $V_g$ is used.  The
  measurement shown in Fig.~\ref{fig4}(a) is made with all voltages
  chosen to optimize the QPC transducer resolution.
\bibitem{footnote:3} The $Q$ plotted here is extracted from spectra of
  the cantilever's thermal motion.  Since these spectra are averaged
  over several minutes, they are susceptible to slow drift in the QPC
  conductance due to charges rearraging in the device.  The plotted
  $Q$ is therefore lower than the actual $Q$, which is more accurately
  measured to be around 2500 using a ``ring-down'' method.
\end{thebibliography}
\end{document}